\documentclass[journal=jacsat,manuscript=article]{achemso}

\usepackage{chemformula} 
\usepackage[T1]{fontenc} 
\usepackage{graphicx}
\usepackage{dcolumn}
\usepackage{bbm}
\usepackage{bm}
\usepackage{hyperref}
\usepackage{makecell} 
\usepackage{scalerel}
\usepackage{enumerate}
\usepackage{ulem}
\usepackage{setspace}

\hypersetup{colorlinks=true,citecolor=blue,urlcolor=blue,linkcolor=blue}

\author{Zhaohua Tian$^{1}$}
\author{Qi Liu $^{1,2}$}
\email{liuqi@stu.pku.edu.cn}
\author{Yu Tian$^{1,2}$}
\author{Ying Gu$^{1,2,3,4,5}$}
\email{ygu@pku.edu.cn}
\affiliation{$^1$State Key Laboratory for Mesoscopic Physics, Department of Physics, Peking University, Beijing 100871, China\\
$^2$Frontiers Science Center for Nano-optoelectronics $\&$  Collaborative Innovation Center of Quantum Matter $\&$ Beijing Academy of Quantum Information Sciences, Peking University, Beijing 100871, China\\
$^3$Collaborative Innovation Center of Extreme Optics, Shanxi University, Taiyuan, Shanxi 030006, China\\
$^4$Peking University Yangtze Delta Institute of Optoelectronics, Nantong 226010, China\\
$^5$Hefei National Laboratory, Hefei 230088, China}

\title{Conversion of photon temporal shape using single gradient metasurface}

\abbreviations{IR,NMR,UV}

\begin{document}







\begin{abstract}
By applying phase modulation across different frequencies, metasurfaces possess the ability to manipulate the temporal dimension of photons at the femtosecond scale. However, there remains a fundamental challenge to shape the single wavepacket at the nanosecond scale by using of metasurfaces. Here, we propose that the single photon temporal shape can be converted through the multi-photon wavepacket interference on a single metasurface. By selecting appropriate input single-photon temporal shapes and metasurfaces beam splitting ratio, controllable  photon shape conversion can be achieved with high fidelity. For examples, photons with an exponentially decaying profile can be shaped into a Gaussian profile; by tuning the relative time delays of input photons, Gaussian-shaped photons can be transformed into exponentially decaying or rising profiles through the same metasurface. The proposed mechanism provides a compact way for solving the temporal shape mismatch issues in quantum networks, facilitating the realization of high-fidelity on-chip quantum information processing.
\end{abstract}


\par The photon temporal shape plays an important role in quantum interference \cite{RempePRL2004}, light-matter interactions \cite{FanPRL2007} and optical quantum information processes \cite{Kimble2008QuanInter} .
Photons with an exponentially rising (ER) shape can be completely loaded into a resonant cavity \cite{LeuchsNJP2013,PRL2014ShengwangDu} as well as fully excite a two-level emitter \cite{PRA2010FanFullInversion,KurtsieferNC2016} . Phototons with time-reversal-symmetric profiles, like Gaussian shapes, have applications in deterministic quantum state transfer \cite{RitterNature2012PQST,LeuchsTimeReverSymmetric} and the implementation of controlled logic gates \cite{PRApplied2020Zou,PRL2020Dirk} . Gaussian shape is also optimal \cite{NielsenPRA2005} for various interference-based quantum information processing \cite{PieteRMP2007} .
However, current solid-state quantum emitters, such as quantum dots, molecules,  and color centers, typically emit photons with an exponentially decaying (ED) profile \cite{MilosNP2016,LodahlPRL2014,RenNatCommun2022} .
To meet the diverse requirements of quantum information tasks, it is essential to have the capability to convert the photons with ED shapes into Gaussian shapes or ER shapes.
Currently, photon temporal shaping mainly relies on the dynamic control of emitters using sophisticated setups \cite{KimbleScience2004,KellerNature2004} , or through the electro-optic effects \cite{HarrisPRL2008,RempeNatPho2009,YuanLPR2022} and nonlinear effects \cite{PRL2011WisemanShapeConversion} of bulk materials.
Nevertheless, these shaping approaches face challenges in integration and scalability, limiting their applications in on-chip quantum information processes.

\par Metasurfaces, a type of two-dimensional metamaterial composed of subwavelength structural elements, can compactly manipulate nearly all dimensions of photons \cite{NanfangRepProgPhys2016Review,QiuNanoLett2021,CapassoScience2022} . 
By applying phase modulation across different frequencies, metasurfaces possess the ability to manipulate the temporal dimension of photons at the femtosecond scale \cite{DivittScience2019MetaFS,SalehiOE2021,ThomasNanoLett2023} .
Theoretically, picosecond pulses with narrower bandwidths can also be manipulated by metasurfaces featuring high-Q modes \cite{HosseinACSPho2020MS,HosseinAPR2023} . 
However, metasurfaces relying on dispersion principles are inherently incapable of shaping single-photon wavepackets at the nanosecond scale, as the phase response remains constant within such a narrow spectral width (typically $<10^{-4}$ nm). 
To address this limitation, new shaping mechanisms are required. 
Here, we propose that by exploiting the parallel beam-splitting capabilities \cite{QiLiuOE2024} , metasurface can be used to achieve temporal shaping of nanosecond-scale photon wavepackets.

\par In this Letter, we propose that high-fidelity conversion of photon temporal shapes can be realized through single phase-gradient metasurface.
The parallel beam-splitting of metasurfaces enables multi-photon wavepacket interference.
By selecting appropriate input single-photon temporal shapes, metasurfaces beam splitting ratio, and time-resolved measurements, controllable photon shape conversion can be realized.
Based on the three-photon wavepacket interference, the photon wavepackets with ED shape can be converted into Gaussian shape with a high fidelity of 99.7\%.  
By adjusting the relative time delay of input photons, Gaussian-shaped photons can also be converted into ED or ER profiles through the same metasurface with fidelity exceeding 97.7\%.
This approach does not rely on the frequency-dependent phase response of metasurfaces, but instead utilizes the multi-photon interference process within an extremely narrow spectral range.
Moreover, this method avoids the need for nonlinear or electro-optic effects of bulk materials, providing a compact and efficient solution to photon temporal shape mismatch issues in integrated quantum networks.

\begin{figure}[!t]
  \includegraphics*[width=75mm]{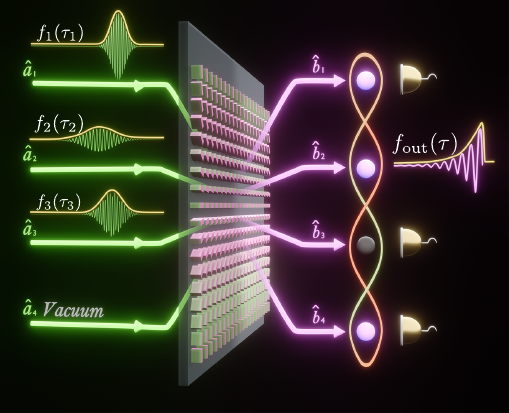}
  \caption{Schematic of photon temporal shaping using single metasurface.
  Through the four-mode beam splitter supported by the metasurface, photon temporal shaping can be achieved via the wavepacket interference of three photons.  }
  \label{fig-1}
\end{figure}

\par The mechanism of photon temporal shaping  based on metasurface is described as follows. As shown in Fig. \ref{fig-1}, multi-photon wavepacket interference can be achieved through a phase-gradient metasurface with parallel beam-splitting properties.  Three photon wavepackets with different temporal shapes are input through separate modes. After passing through the metasurface, they become temporally entangled. By detecting two photons via joint time-resolved measurements (TRMs), the temporal shape of the remaining single photon becomes a superposition of those of input photons. Through appropriate selection of input photon temporal profiles, metasurface beam splitting ratios, and time-resolved measurements, controllable  photon shape conversion can be realized with high fidelity. This approach does not rely on the frequency-dependent phase response of metasurfaces \cite{DivittScience2019MetaFS,SalehiOE2021,ThomasNanoLett2023,HosseinACSPho2020MS,HosseinAPR2023} but instead utilizes the multi-photon wavepacket interference process, which is crucial for enabling the metasurface to shape photon wavepackets at the nanosecond scale.

\par Consider the gradient metasurface with parallel beam-splitting properties shown in Fig. \ref{fig-1}. According to the polarization-dependent beam splitting  property, the metasurface can support a series of parallel two-mode beam splitting processes among adjacent diffraction orders with circular polarization \cite{QiLiuOE2024} . Here, by changing the circular polarization basis into a linear one, the two adjacent beam splitting processes will be coupled, forming an equivalent four-mode beam splitter (BS) \cite{TianSupp} . The annihilation of photons in the $n$-th input (output) mode is denoted by the continuous annihilation operator $\hat{a}_{n}(\tau)$ [$\hat{b}_{n}(\tau)$] \cite{LoudonJoAB1989} . To illustrate the characteristics of the four-mode beam splitting, modes $\hat{b}_{2}$ and $\hat{b}_{3}$ are depicted as two separate paths in the schematic of Fig. \ref{fig-1}. Actually, modes $\hat{b}_{2}$ and $\hat{b}_{3}$ correspond to two linear polarization modes in the same path. The linear relation between the input and output operators can be described through the transformation matrix $\bm{\mathrm{S}}$ as follows

\begin{equation}\label{eq:transMat}
\left[\begin{array}{c}
 \hat{b}_{1}(\tau) \\
 \hat{b}_{2}(\tau) \\
 \hat{b}_{3}(\tau) \\
 \hat{b}_{4}(\tau) 
\end{array} \right]
=\left[\begin{array}{cccc}
  s_{t} & \frac{i }{\sqrt{2}}s_{r} & \frac{-1}{\sqrt{2}}s_{r} & 0 \\
  i s_{r} & \frac{1}{\sqrt{2}}s_{t} & \frac{i }{\sqrt{2}}s_{t} & 0 \\
  0 & \frac{1}{\sqrt{2}}s_{t} & \frac{-i }{\sqrt{2}}s_{t} & i s_{r} \\
  0 & \frac{i }{\sqrt{2}}s_{r} & \frac{1}{\sqrt{2}}s_{r} & s_{t}
\end{array}\right]
\left[\begin{array}{c}
 \hat{a}_{1}(\tau) \\
 \hat{a}_{2}(\tau) \\
 \hat{a}_{3}(\tau) \\
 \hat{a}_{4}(\tau) 
\end{array} \right],
\end{equation}
where $s_{t},s_{r}$ are the co-polarization and cross-polarization conversion coefficients for input photons with circular polarization ($s_{t}^2+s_{r}^2=1$), which are determined by the size and the material of meta units of metasurface \cite{QiLiuOE2024} . The unitary transformation  in Eq.  \eqref{eq:transMat} maintains the communtation relations of input and output operators $[\hat{a}_{m}(\tau),\hat{a}_{n}^{\dagger}(\tau')]=\delta_{m,n}\delta(\tau-\tau'), [\hat{b}_{m}(\tau),\hat{b}_{n}^{\dagger}(\tau')]=\delta_{m,n}\delta(\tau-\tau')$. The effective Hamiltonian $\hat{H}$ for the four-mode beam-splitting process can be obtained according to the transformation matrix $\bm{\mathrm{S}}$ \cite{TianSupp} . Then the transformation from input state into output state can be described as $|\psi\rangle_{\text{out}}=\hat{U}|\psi\rangle_{\text{in}}$, where $\hat{U}=e^{-i\hat{H}t/\hbar}$ is the time evolution operator. This four-mode BS can facilitate multi-photon interference, enabling photon temporal shaping. 
\par Next, we consider three photon wavepackets with different temporal shapes  input through the first three input modes of the metasurface. The input state is represented as
\begin{equation}\label{eq:psiin}
|\psi\rangle_{\text{in}}=\prod_{n=1}^{3}\int f_{n}(\tau_{n})\hat{a}_{n}^{\dagger}(\tau_{n})\mathrm{d}\tau_{n}|0,0,0,0\rangle,
\end{equation}
where  $f_{n}(\tau_{n})$ denotes the photon temporal shape in the $n$-th input mode, normalized by $\int |f_{n}(\tau_{n})|d\tau_{n}=1$. Here $|0,0,0,0\rangle$ represents the vacuum state. After passing through the metasurface, the output state can be obtained through $|\psi\rangle_{\text{out}}=\hat{U}|\psi\rangle_{\text{in}}$ which takes the following exact form:

\begin{equation}\label{eq:psiout}
  \begin{aligned}
    & |\psi\rangle_{\text {out }}=\prod_{n=1}^3 \int f_n\left(\tau_n\right)\left[\sum_{m=1}^4 S_{m, n} \hat{b}_n^{\dagger}\left(\tau_n\right)\right]\mathrm{d}\tau_{n}|0, 0,0, 0\rangle \\
    & =\sum_{j=1}^{20} \sqrt{P_{j}}|m_{1},m_{2},m_{3},m_{4}\rangle_{j}.
  \end{aligned}
\end{equation}
Here $|m_{1},m_{2},m_{3},m_{4}\rangle_{j}$ represents a specific photon number distribution with $m_{n}$ photons in the $n$-th output mode, satisfying $\sum_{n=1}^{4}m_{n}=3$. $P_{j}$ denotes the probability associated with the component $|m_{1},m_{2},m_{3},m_{4}\rangle_{j}$, satisfying $\sum_{j=1}^{20}P_{j}=1$. The exact form of each output component is given by:
\begin{equation}
  |m_{1},m_{2},m_{3},m_{4}\rangle_{j}=\iiint F_{j}(\mathcal{T}) \hat{D}_{j}^{\dagger}(\mathcal{T}) \mathrm{d} \mathcal{T}|0, 0,0, 0\rangle,
\end{equation}
where
\begin{equation}\label{eq:fmalpha}
  \begin{aligned}
    & F_{j}(\mathcal{T})=\frac{\sum_{n_{k, l}}^{\mathrm{permu}} \prod_{k=1}^{N=4}\sum_{\tau_{k, l}}^{\text {permu }} \frac{\prod_{l=1}^{m_k} S_{k, n_{k, l}}  f_{n_{k, l}}\left(\tau_{k, l}\right)}{\sqrt{m_{k}!}}}{\sqrt{P_{j}}}. \\
    &
  \end{aligned}
\end{equation}
Here $\prod_{l=1}^{m_{k}}\{\}=1$ when $m_{k}=0$. $\hat{D}_{j}^{\dagger}(\mathcal{T})=\prod_{k=1}^{4}\frac{\prod_{l=1}^{m_{k}}\hat{b}_{k}^{\dagger}(\tau_{k,l})}{\sqrt{m_{k}!}}$ is the joint creation operator of component $|m_{1},m_{2},m_{3},m_{4}\rangle_{j}$ and $F_{j}(\mathcal{T})$ is the corresponding wavefunction. $\mathcal{T}=[\tau_{1,1},\tau_{1,2},...]^{\mathrm{T}}$ denotes the temporal variables of the multi-photon wavefunction and $\tau_{k,l}$ corresponding to the $l$-th photon in the output mode $\hat{b}_{k}$. 
In Eq. \eqref{eq:fmalpha}, $\sum_{\tau_{k,l}}^{\text{permu}}$ means the exchange of indices and summation required when multiple photons occupy the same output mode $\hat{b}_{k}$; $\sum_{n_{k, l}}^{\mathrm{permu}}$ represents the summation over all possible photon transformations leading to the same outcome  $|m_{1},m_{2},m_{3},m_{4}\rangle_{j}$. The detailed expressions of the wavefunction are provided in the supplementary materials \cite{TianSupp} . It can be seen from Eq. \eqref{eq:fmalpha} that  the wavefunctions are inseparable, i.e., the photons are temporally entangled \cite{PRX2015QuanInterPhilip,PRA2019HOMSilberhorn,JOSABTZH2024} due the multi-photon wavepacket interference. Such temporal entanglement through metasurface is crucial for enabling single photon temporal shape conversion.

\par Then, we perform joint TRMs on the temporally entangled photons to achieve shape conversion. By detecting two of the output photons at times $t_{\text{dec},1}$ and $t_{\text{dec},2}$ using photon number resolved detectors (PNRDs), a specific component $|m_{1},m_{2},m_{3},m_{4}\rangle_{j}$ can be probabilistically selected. The temporal shape of the remaining single photon is given by \cite{TianSupp}
\begin{equation}
\begin{aligned}
f_{\text {out }}(\tau) & =F_j\left(\tau| t_{\text {dec}, 1}, t_{\text {dec}, 2}\right) \\
& =\sum_{n=1}^3 \xi_n\left(s_{t}, t_{\text {dec}, 1}, t_{\text {dec}, 2}\right) f_n(\tau).
\end{aligned}
\end{equation}
One sees that the output photon temporal shape $f_{\text {out }}(\tau)$  is a coherent superposition of the three input photons $f_n(\tau)$, with the superposition coefficients $\xi_n$ determined by the detection time ($t_{\text {dec }, 1}, t_{\text {dec }, 2}$) and the structure of the metasurface itself ($s_{t}$).
Therefore, by selecting appropriate input single-photon temporal shapes, metasurface beam splitting ratio, and time-resolved measurements, controllable and high-fidelity photon shape conversion can be achieved.

\begin{figure}[!t]
  \includegraphics*[width=86mm]{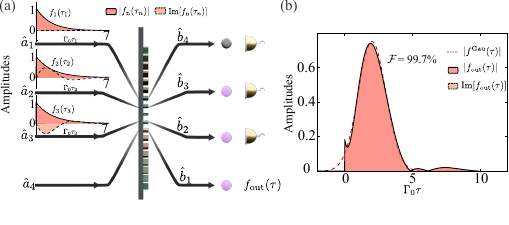}
  \caption{ Conversion of photons from ED shape into Gaussian shape. (a) Temporal shaping scheme based on the output component $|1,1,1,0\rangle$. Three PNRDs are placed at the out modes $\hat{b}_{2}$ to $\hat{b}_{4}$. (b) The temporal profile of the resulting single photon at the output mode $\hat{b}_{1}$.}
  \label{fig-2}
\end{figure}

\begin{figure*}[!t]
  \includegraphics*[width=155mm]{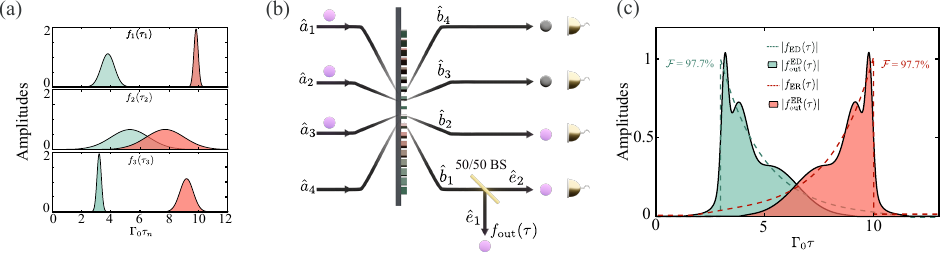}
  \caption{ Conversion of photons from Gaussian shape into ED and ER shapes through the same piece of metasurface. (a) The temporal profile of three input Gaussian-shaped photons for converting into ER shape (orange color) and ED shape (green color);  (b) Schematic of temporal shaping based on the output component  $|2,1,0,0\rangle$. Four PNRDs are required and an extra 50/50 BS should be placed at the out mode $\hat{b}_{1}$ of the metasurface. (c) The temporal profiles of the converted ER- and ED-shaped photons in out mode $\hat{b}_{1}$ of 50/50 BS.}
  \label{fig-3}
\end{figure*}

\par Based on the temporal shaping mechanism mentioned above, we first demonstrate the conversion of photon wavepacket from an ED to a Gaussian shape. 
Three ED-shaped photons denoted as $f_{n}^{\text{ED}}(\tau )=\sqrt{\Gamma _n} \Theta (\tau) e^{-i\left(\omega _n-i\frac{\Gamma _n}{2}\right)\tau}$ are input through modes $\hat{a}_{1}, \hat{a}_{2}, \hat{a}_{3}$  with $\Theta(\tau)$ being the Heaviside step function [Fig. \ref{fig-2}(a)]. Here the co-polarization conversion coefficient of metasurface is $s_{t}=0.801$, specific parameters of the designed metasurface can be found in the supporting information \cite{TianSupp} . The linewidths and frequencies of input photons  normalized by the linewidth of the target photon $\Gamma_{0}$ are given by
$(\Gamma_{1},\Gamma_{2},\Gamma_{3})/\Gamma_{0}=(1.42, 1.27, 1.27)$, $(\Delta\omega_{1},\Delta\omega_{2},\Delta\omega_{3})/\Gamma_{0}=(0, -1.19, 1.19)$.
Here $\Delta\omega_{n}=\omega_{n}-\omega_{0}$ and $\omega_{0}$ is the frequency of the target photon, throughout this paper we set $\omega_{0}=0$ for convenience. Three photon wavepackets interfere through the metasurface, then by using three PNRDs, the output state component $|1,1,1,0\rangle$ in Eq. \eqref{eq:psiout} will be selected for shaping with a selection probability of $P_\text{sel}=P_{|1,1,1,0\rangle}=8.86\%$.
When a single-photon is detected at output modes $\hat{b}_{2}$ and $\hat{b}_{3}$ at times $\Gamma_{0} t_{\text{dec},1}=0.906,\Gamma_{0} t_{\text{dec},2}=1.42$, respectively, and no photon is detected at output mode $\hat{b}_{4}$, the final photon will emerge from mode $\hat{b}_{1}$.
The temporal profile of the remaining output photon $f_{\text{out}}(\tau)$ is nearly overlapped with the target Gaussian-shape $f^{\text{Gau}}(\tau )=\sqrt{\frac{\Gamma_{0} }{\sqrt{\pi }}} e^{-\frac{1}{2} \Gamma _{0}^2 \left(\tau -\tau _{0}\right)^2}$\quad ($\Gamma_{0}\tau _{0}=2$) with high fidelity of $\mathcal{F}=\left| \int f_{\text{out}}^{*}(\tau ) f^{\text{Gau}}(\tau) d\tau \right|=99.7\%$.
Thus, through the interference process of the metasurface, the photon with an ED shape is converted into a Gaussian shape. The generated time-reversal-symmetric Gaussian-shaped photon will have potential applications in various interference-based \cite{NielsenPRA2005} and matter-based quantum information processes \cite{RitterNature2012PQST,PRL2020Dirk,PRApplied2020Zou,LeuchsTimeReverSymmetric} .

\par We further demonstrate that Gaussian-shaped photons can be converted into an ER shape and an ED shape with high fidelity. 
Three Gaussian-shaped photons with the same frequency, as shown in orange color of Fig. \ref{fig-3}(a), are input through modes $\hat{a}_{1}, \hat{a}_{2}, \hat{a}_{3}$, where their linewidths and time delays are
$(\Gamma_{1},\Gamma_{2},\Gamma_{3})/\Gamma_{0}=(6.73, 0.771, 2.18)$, $(\tau_{0,1},\tau_{0,2},\tau_{0,3})\Gamma_{0}=(9.80, 7.71, 9.21)$, and the co-polarization conversion coefficient of the metasurface is $s_{t}=0.782$.
Through wavepacket interference on the metasurface, the output state component $|2,1,0,0\rangle$ in Eq. \eqref{eq:psiout} will be selected for shaping. Four PNRDs are required, along with an additional 50/50 BS at the output mode $\hat{b}_{1}$  [Fig. \ref{fig-3}(b)]. The probability of selecting component $|2,1,0,0\rangle$ is $P_{\text{sel}}=P_{|2,1,0,0\rangle}/2=4.2\%$. 
By detecting a single photon at the output mode $\hat{e}_{2}$ of  BS with $\Gamma_{0} t_{\text{dec},1}=9.68$ and output mode $\hat{b}_{2}$ of the metasurface with $\Gamma_{0} t_{\text{dec},2}=10.2$, the remaining single photon  emerges from the output mode $\hat{e}_{1}$ of the BS. 
The temporal profile of the resulting single photon $f_{\text{out}}^{\text{ER}}(\tau )$ is very similar to the target ER shape $f^{\text{ER}}(\tau )=\sqrt{\Gamma _{0}}e^{\Gamma_{0}(\tau -\tau _{0})/2}\Theta \left(\tau _{0}-\tau \right)$ with $\Gamma_{0}\tau _{0}=10$, which is shown as orange color in Fig. \ref{fig-3}(c).   
As can be seen, through the interference process of the metasurface, the photon with an Gaussian shape is converted into an ER shape with a fidelity of $\mathcal{F}=97.7\%$. 
The ER-shaped photons generated by this conversion process may have important applications in efficient excitation of quantum emitters \cite{PRA2010FanFullInversion,LeuchsPerfectExcitFreeSpace,KurtsieferNC2016} and storage of photonic qubits \cite{PRL2014ShengwangDu,LeuchsNJP2013} .

\par Using the same metasurface with $s_{t}=0.782$, if only changing the input mode and time delay of input photon wavepackets, it is also possible to convert Gaussian-shaped photons into an ED shape.
As shown in Fig. \ref{fig-3}(a), three input photon wavepackets in orange color are replaced by those in green color with 
$(\Gamma_{1},\Gamma_{2},\Gamma_{3})/\Gamma_{0}=(2.18, 0.771, 6.73)$, $(\tau_{0,1},\tau_{0,2},\tau_{0,3})\Gamma_{0}=(3.79, 5.29, 3.20)$.
With the same joint TRMs, the detection time now becomes $\Gamma_{0}t_{\text{dec},1}=3.44, \Gamma_{0}t_{\text{dec},2}=2.87$ and the probability of selecting component $|2,1,0,0\rangle$ is $P_{\text{sel}}=P_{|2,1,0,0\rangle}/2=5.65\%$. 
It can be seen from green color in Fig. \ref{fig-3}(c) that through the interference process of the metasurface, the Gaussian shaped photons can also be converted into an ED shape $f_{\text{out}}^{\text{ED}}(\tau)$  with a fidelity of  $97.7\%$ in output mode $\hat{e}_{1}$. 
According to Fig. \ref{fig-2}(b) and Fig. \ref{fig-3}(c), the interconversion between Gaussian-shaped and ED-shaped photons can be achieved, which will contribute to addressing the issue of photon shape mismatch in quantum networks \cite{PRL2011WisemanShapeConversion}.

\begin{table}[!b] 

\caption{Conversion of single photon from ED shape into Gaussian shape based on different output componenents and detection modes.}

\begin{tabular}{ccccc}
\Xhline{0.8pt}
$s_{t}$ & $|m_{1},m_{2},m_{3},m_{4}\rangle$ & $ P_{\text{sel}}(\%)$\quad   & Output   & $\quad \mathcal{F}(\%)$ \\ \Xhline{0.8pt}
0.801        & $|1,1,1,0\rangle$                          & 8.86    & $|1,0,0,0\rangle$ & 99.7        \\ 
0.615        & $|1,1,1,0\rangle$                          & 5.10    & $|0,1,0,0\rangle$ & 99.7       \\ 
0.994        & $|1,1,1,0\rangle$                          & 28.0      & $|0,0,1,0\rangle$ & 98.3       \\ 
0.621        & $|2,1,0,0\rangle$                          & 4.98      & $|1,0,0,0\rangle$ & 99.6        \\ 
0.726        & $|1,2,0,0\rangle$                          & 5.45    & $|1,0,0,0\rangle$ & 99.5        \\ \Xhline{0.8pt}
\end{tabular}
\label{table1}
\end{table}

\par  We now examine the differences in shaping results by selecting different output components. We take the conversion from a Gaussian shape to an ED shape as a specific example.  As shown in the first three rows of Tab. \ref{table1}, when selecting the component $|1,1,1,0\rangle$ for shaping, high fidelity of 99.7\% can still be obtained when the remaining photon exits from modes $\hat{b}_{2}$. By slightly relaxing the fidelity requirement to 98.3\%, the selection probability can be significantly improved to $28\%$ when output from mode $\hat{b}_{3}$. 
High conversion fidelity is also attainable with the components $|2,1,0,0\rangle$ and $|1,2,0,0\rangle$, but an extra PNRD and a 50/50 BS are required (Similar to the setup in Fig. \ref{fig-3}(b)). We also examined the shaping process for inputting photons through the modes $\hat{b}_{1}, \hat{b}_{2}, \hat{b}_{4}$ and found that a high fidelity of 98.3\% can still be obtained \cite{TianSupp}. In short, flexible shape conversions can be realized through metasurface, which are not limited to certain input, output, and detection methods.
\par
\begin{figure}[!t]
  \includegraphics[width=86mm]{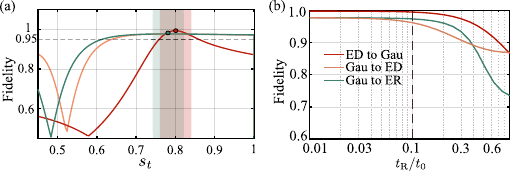}
  \caption{ Shaping fidelity as a function of  (a) copolarization conversion coefficient $s_{t}$ of the metasurface and (b) the detector's temporal resolution $t_{\text{R}}/t_{0}$ for three different shaping processes. }
  \label{fig-4}
\end{figure}

 The proposed shaping mechanism depends on the beam splitting capability of metasurface, the influence of the co-polarization conversion coefficients $s_{t}$ on the shaping fidelity is depicted in Fig. \ref{fig-4}(a).
For the three conversion processes, the fidelities slightly decrease as the coefficients $s_{t}$ deviate from their optimal values. However, a high fidelity of 95\% can still be obtained if the relative error of co-polarization conversion coefficient $s_{t}$ is within  $\pm$5 \% [the shaded green and red regions of Fig. \ref{fig-4}(b)], which is feasible in experiments \cite{GuixinNanoLett2023,PRL2017Capasso} .

\par The shaping fidelity will also depend on the resolution of TRMs. Assuming we can only confirm the detection of photons within a certain time interval  $(t_{\text{dec}}-t_{\text{R}}/2,t_{\text{dec}}+t_{\text{R}}/2)$, where $t_{\text{R}}$ denotes the resolution time of the detector.  
The influences of the ratio $t_{\text{R}}/t_{0}$ on the fidelities are depicted in Fig. \ref{fig-4}(b) where $t_{0}=1/\Gamma_{0}$ denotes the temporal width of target photon (see supporting information for details \cite{TianSupp}) .
It can be seen that as the resolution decreases ($t_{\text{R}}$ increases), the fidelities decrease too.
Nevertheless, when $t_{\text{R}}/t_{0}<0.1$, the fidelities remain nearly unaffected. Such precise TRMs have been experimentally demonstrated \cite{RempePRL2004,PanPRL2018,AnthonyPRL2024} . Therefore, it's promising to achieve high fidelity shape conversion based on current detection technology. 
\par In summary, we have proposed a wavepacket shaping mechanism based on the multi-photon wavepacket interference, which addresses the critical challenge of controlling nanosecond-scale photon wavepackets with metasurfaces. 
By selecting appropriate input single-photon temporal shapes, metasurfaces beam splitting ratio, and time-resolved measurements, controllable and high fidelity photon shape conversion have been realized. Through three-photon wavepacket interference, we demonstrated the interconversion between ED and Gaussian shapes, with Gaussian-shaped photons also convertible to an ER shape.
These conversion processes hold potential applications in deterministic quantum state transfer and quantum-controlled logic gates. The proposed shaping mechanism using metasurface provides a compact solution to the temporal shape mismatch issues in quantum networks, facilitating the realization of high-fidelity on-chip quantum information processing.

\begin{acknowledgement}
This work is supported by the National Natural Science Foundation of China under Grants No. 12474370 and NO. 11974032 and the Innovation Program for Quantum Science and Technology under Grant No. 2021ZD0301500.
\end{acknowledgement}

\begin{suppinfo}
The supporting information includes:
Four-mode beam splitting through the metasurface;
Analytical expressions of the temporal shaping;
Shape conversions based on different output components;
Influence of the detector’s resolution on the fidelity.
%

\end{suppinfo}


\providecommand{\latin}[1]{#1}
\makeatletter
\providecommand{\doi}
  {\begingroup\let\do\@makeother\dospecials
  \catcode`\{=1 \catcode`\}=2 \doi@aux}
\providecommand{\doi@aux}[1]{\endgroup\texttt{#1}}
\makeatother
\providecommand*\mcitethebibliography{\thebibliography}
\csname @ifundefined\endcsname{endmcitethebibliography}
  {\let\endmcitethebibliography\endthebibliography}{}
\begin{mcitethebibliography}{42}
\providecommand*\natexlab[1]{#1}
\providecommand*\mciteSetBstSublistMode[1]{}
\providecommand*\mciteSetBstMaxWidthForm[2]{}
\providecommand*\mciteBstWouldAddEndPuncttrue
  {\def\EndOfBibitem{\unskip.}}
\providecommand*\mciteBstWouldAddEndPunctfalse
  {\let\EndOfBibitem\relax}
\providecommand*\mciteSetBstMidEndSepPunct[3]{}
\providecommand*\mciteSetBstSublistLabelBeginEnd[3]{}
\providecommand*\EndOfBibitem{}
\mciteSetBstSublistMode{f}
\mciteSetBstMaxWidthForm{subitem}{(\alph{mcitesubitemcount})}
\mciteSetBstSublistLabelBeginEnd
  {\mcitemaxwidthsubitemform\space}
  {\relax}
  {\relax}

\bibitem[Legero \latin{et~al.}(2004)Legero, Wilk, Hennrich, Rempe, and
  Kuhn]{RempePRL2004}
Legero,~T.; Wilk,~T.; Hennrich,~M.; Rempe,~G.; Kuhn,~A. Quantum Beat of Two
  Single Photons. \emph{Phys. Rev. Lett.} \textbf{2004}, \emph{93},
  070503\relax
\mciteBstWouldAddEndPuncttrue
\mciteSetBstMidEndSepPunct{\mcitedefaultmidpunct}
{\mcitedefaultendpunct}{\mcitedefaultseppunct}\relax
\EndOfBibitem
\bibitem[Shen and Fan(2007)Shen, and Fan]{FanPRL2007}
Shen,~J.-T.; Fan,~S. Strongly Correlated Two-Photon Transport in a
  One-Dimensional Waveguide Coupled to a Two-Level System. \emph{Phys. Rev.
  Lett.} \textbf{2007}, \emph{98}, 153003\relax
\mciteBstWouldAddEndPuncttrue
\mciteSetBstMidEndSepPunct{\mcitedefaultmidpunct}
{\mcitedefaultendpunct}{\mcitedefaultseppunct}\relax
\EndOfBibitem
\bibitem[Kimble(2008)]{Kimble2008QuanInter}
Kimble,~H.~J. The quantum internet. \emph{Nature} \textbf{2008}, \emph{453},
  1023--1030\relax
\mciteBstWouldAddEndPuncttrue
\mciteSetBstMidEndSepPunct{\mcitedefaultmidpunct}
{\mcitedefaultendpunct}{\mcitedefaultseppunct}\relax
\EndOfBibitem
\bibitem[Bader \latin{et~al.}(2013)Bader, Heugel, Chekhov, Sondermann, and
  Leuchs]{LeuchsNJP2013}
Bader,~M.; Heugel,~S.; Chekhov,~A.~L.; Sondermann,~M.; Leuchs,~G. Efficient
  coupling to an optical resonator by exploiting time-reversal symmetry.
  \emph{New J. Phys.} \textbf{2013}, \emph{15}, 123008\relax
\mciteBstWouldAddEndPuncttrue
\mciteSetBstMidEndSepPunct{\mcitedefaultmidpunct}
{\mcitedefaultendpunct}{\mcitedefaultseppunct}\relax
\EndOfBibitem
\bibitem[Liu \latin{et~al.}(2014)Liu, Sun, Zhao, Zhang, Loy, and
  Du]{PRL2014ShengwangDu}
Liu,~C.; Sun,~Y.; Zhao,~L.; Zhang,~S.; Loy,~M. M.~T.; Du,~S. Efficiently
  Loading a Single Photon into a Single-Sided Fabry-Perot Cavity. \emph{Phys.
  Rev. Lett.} \textbf{2014}, \emph{113}, 133601\relax
\mciteBstWouldAddEndPuncttrue
\mciteSetBstMidEndSepPunct{\mcitedefaultmidpunct}
{\mcitedefaultendpunct}{\mcitedefaultseppunct}\relax
\EndOfBibitem
\bibitem[Rephaeli \latin{et~al.}(2010)Rephaeli, Shen, and
  Fan]{PRA2010FanFullInversion}
Rephaeli,~E.; Shen,~J.-T.; Fan,~S. Full inversion of a two-level atom with a
  single-photon pulse in one-dimensional geometries. \emph{Phys. Rev. A}
  \textbf{2010}, \emph{82}, 033804\relax
\mciteBstWouldAddEndPuncttrue
\mciteSetBstMidEndSepPunct{\mcitedefaultmidpunct}
{\mcitedefaultendpunct}{\mcitedefaultseppunct}\relax
\EndOfBibitem
\bibitem[Leong \latin{et~al.}(2016)Leong, Seidler, Steiner, Cer{\`e}, and
  Kurtsiefer]{KurtsieferNC2016}
Leong,~V.; Seidler,~M.~A.; Steiner,~M.; Cer{\`e},~A.; Kurtsiefer,~C.
  Time-resolved scattering of a single photon by a single atom. \emph{Nat.
  Commun.} \textbf{2016}, \emph{7}, 13716\relax
\mciteBstWouldAddEndPuncttrue
\mciteSetBstMidEndSepPunct{\mcitedefaultmidpunct}
{\mcitedefaultendpunct}{\mcitedefaultseppunct}\relax
\EndOfBibitem
\bibitem[Ritter \latin{et~al.}(2012)Ritter, N{\"o}lleke, Hahn, Reiserer,
  Neuzner, Uphoff, M{\"u}cke, Figueroa, Bochmann, and
  Rempe]{RitterNature2012PQST}
Ritter,~S.; N{\"o}lleke,~C.; Hahn,~C.; Reiserer,~A.; Neuzner,~A.; Uphoff,~M.;
  M{\"u}cke,~M.; Figueroa,~E.; Bochmann,~J.; Rempe,~G. An elementary quantum
  network of single atoms in optical cavities. \emph{Nature} \textbf{2012},
  \emph{484}, 195--200\relax
\mciteBstWouldAddEndPuncttrue
\mciteSetBstMidEndSepPunct{\mcitedefaultmidpunct}
{\mcitedefaultendpunct}{\mcitedefaultseppunct}\relax
\EndOfBibitem
\bibitem[Trautmann \latin{et~al.}(2015)Trautmann, Alber, Agarwal, and
  Leuchs]{LeuchsTimeReverSymmetric}
Trautmann,~N.; Alber,~G.; Agarwal,~G.~S.; Leuchs,~G. Time-Reversal-Symmetric
  Single-Photon Wave Packets for Free-Space Quantum Communication. \emph{Phys.
  Rev. Lett.} \textbf{2015}, \emph{114}, 173601\relax
\mciteBstWouldAddEndPuncttrue
\mciteSetBstMidEndSepPunct{\mcitedefaultmidpunct}
{\mcitedefaultendpunct}{\mcitedefaultseppunct}\relax
\EndOfBibitem
\bibitem[Li \latin{et~al.}(2020)Li, Zhang, Tang, Dong, Guo, and
  Zou]{PRApplied2020Zou}
Li,~M.; Zhang,~Y.-L.; Tang,~H.~X.; Dong,~C.-H.; Guo,~G.-C.; Zou,~C.-L.
  Photon-Photon Quantum Phase Gate in a Photonic Molecule with
  ${\ensuremath{\chi}}^{(2)}$ Nonlinearity. \emph{Phys. Rev. Appl.}
  \textbf{2020}, \emph{13}, 044013\relax
\mciteBstWouldAddEndPuncttrue
\mciteSetBstMidEndSepPunct{\mcitedefaultmidpunct}
{\mcitedefaultendpunct}{\mcitedefaultseppunct}\relax
\EndOfBibitem
\bibitem[Heuck \latin{et~al.}(2020)Heuck, Jacobs, and Englund]{PRL2020Dirk}
Heuck,~M.; Jacobs,~K.; Englund,~D.~R. Controlled-Phase Gate Using Dynamically
  Coupled Cavities and Optical Nonlinearities. \emph{Phys. Rev. Lett.}
  \textbf{2020}, \emph{124}, 160501\relax
\mciteBstWouldAddEndPuncttrue
\mciteSetBstMidEndSepPunct{\mcitedefaultmidpunct}
{\mcitedefaultendpunct}{\mcitedefaultseppunct}\relax
\EndOfBibitem
\bibitem[Rohde \latin{et~al.}(2005)Rohde, Ralph, and Nielsen]{NielsenPRA2005}
Rohde,~P.~P.; Ralph,~T.~C.; Nielsen,~M.~A. Optimal photons for
  quantum-information processing. \emph{Phys. Rev. A} \textbf{2005}, \emph{72},
  052332\relax
\mciteBstWouldAddEndPuncttrue
\mciteSetBstMidEndSepPunct{\mcitedefaultmidpunct}
{\mcitedefaultendpunct}{\mcitedefaultseppunct}\relax
\EndOfBibitem
\bibitem[Kok \latin{et~al.}(2007)Kok, Munro, Nemoto, Ralph, Dowling, and
  Milburn]{PieteRMP2007}
Kok,~P.; Munro,~W.~J.; Nemoto,~K.; Ralph,~T.~C.; Dowling,~J.~P.; Milburn,~G.~J.
  Linear optical quantum computing with photonic qubits. \emph{Rev. Mod. Phys.}
  \textbf{2007}, \emph{79}, 135--174\relax
\mciteBstWouldAddEndPuncttrue
\mciteSetBstMidEndSepPunct{\mcitedefaultmidpunct}
{\mcitedefaultendpunct}{\mcitedefaultseppunct}\relax
\EndOfBibitem
\bibitem[Aharonovich \latin{et~al.}(2016)Aharonovich, Englund, and
  Toth]{MilosNP2016}
Aharonovich,~I.; Englund,~D.; Toth,~M. Solid-state single-photon emitters.
  \emph{Nat. Photonics} \textbf{2016}, \emph{10}, 631--641\relax
\mciteBstWouldAddEndPuncttrue
\mciteSetBstMidEndSepPunct{\mcitedefaultmidpunct}
{\mcitedefaultendpunct}{\mcitedefaultseppunct}\relax
\EndOfBibitem
\bibitem[Arcari \latin{et~al.}(2014)Arcari, S\"ollner, Javadi, Lindskov~Hansen,
  Mahmoodian, Liu, Thyrrestrup, Lee, Song, Stobbe, and Lodahl]{LodahlPRL2014}
Arcari,~M.; S\"ollner,~I.; Javadi,~A.; Lindskov~Hansen,~S.; Mahmoodian,~S.;
  Liu,~J.; Thyrrestrup,~H.; Lee,~E.~H.; Song,~J.~D.; Stobbe,~S.; Lodahl,~P.
  Near-Unity Coupling Efficiency of a Quantum Emitter to a Photonic Crystal
  Waveguide. \emph{Phys. Rev. Lett.} \textbf{2014}, \emph{113}, 093603\relax
\mciteBstWouldAddEndPuncttrue
\mciteSetBstMidEndSepPunct{\mcitedefaultmidpunct}
{\mcitedefaultendpunct}{\mcitedefaultseppunct}\relax
\EndOfBibitem
\bibitem[Ren \latin{et~al.}(2022)Ren, Wei, Liu, Lin, Tian, Huang, Tang, Shi,
  and Chen]{RenNatCommun2022}
Ren,~P.; Wei,~S.; Liu,~W.; Lin,~S.; Tian,~Z.; Huang,~T.; Tang,~J.; Shi,~Y.;
  Chen,~X.-W. Photonic-circuited resonance fluorescence of single molecules
  with an ultrastable lifetime-limited transition. \emph{Nat. Commun.}
  \textbf{2022}, \emph{13}, 3982\relax
\mciteBstWouldAddEndPuncttrue
\mciteSetBstMidEndSepPunct{\mcitedefaultmidpunct}
{\mcitedefaultendpunct}{\mcitedefaultseppunct}\relax
\EndOfBibitem
\bibitem[McKeever \latin{et~al.}(2004)McKeever, Boca, Boozer, Miller, Buck,
  Kuzmich, and Kimble]{KimbleScience2004}
McKeever,~J.; Boca,~A.; Boozer,~A.~D.; Miller,~R.; Buck,~J.~R.; Kuzmich,~A.;
  Kimble,~H.~J. Deterministic Generation of Single Photons from One Atom
  Trapped in a Cavity. \emph{Science} \textbf{2004}, \emph{303},
  1992--1994\relax
\mciteBstWouldAddEndPuncttrue
\mciteSetBstMidEndSepPunct{\mcitedefaultmidpunct}
{\mcitedefaultendpunct}{\mcitedefaultseppunct}\relax
\EndOfBibitem
\bibitem[Keller \latin{et~al.}(2004)Keller, Lange, Hayasaka, Lange, and
  Walther]{KellerNature2004}
Keller,~M.; Lange,~B.; Hayasaka,~K.; Lange,~W.; Walther,~H. Continuous
  generation of single photons with controlled waveform in an ion-trap cavity
  system. \emph{Nature} \textbf{2004}, \emph{431}, 1075--1078\relax
\mciteBstWouldAddEndPuncttrue
\mciteSetBstMidEndSepPunct{\mcitedefaultmidpunct}
{\mcitedefaultendpunct}{\mcitedefaultseppunct}\relax
\EndOfBibitem
\bibitem[Kolchin \latin{et~al.}(2008)Kolchin, Belthangady, Du, Yin, and
  Harris]{HarrisPRL2008}
Kolchin,~P.; Belthangady,~C.; Du,~S.; Yin,~G.~Y.; Harris,~S.~E. Electro-Optic
  Modulation of Single Photons. \emph{Phys. Rev. Lett.} \textbf{2008},
  \emph{101}, 103601\relax
\mciteBstWouldAddEndPuncttrue
\mciteSetBstMidEndSepPunct{\mcitedefaultmidpunct}
{\mcitedefaultendpunct}{\mcitedefaultseppunct}\relax
\EndOfBibitem
\bibitem[Specht \latin{et~al.}(2009)Specht, Bochmann, M{\"u}cke, Weber,
  Figueroa, Moehring, and Rempe]{RempeNatPho2009}
Specht,~H.~P.; Bochmann,~J.; M{\"u}cke,~M.; Weber,~B.; Figueroa,~E.;
  Moehring,~D.~L.; Rempe,~G. Phase shaping of single-photon wave packets.
  \emph{Nat. Photonics} \textbf{2009}, \emph{3}, 469--472\relax
\mciteBstWouldAddEndPuncttrue
\mciteSetBstMidEndSepPunct{\mcitedefaultmidpunct}
{\mcitedefaultendpunct}{\mcitedefaultseppunct}\relax
\EndOfBibitem
\bibitem[Li \latin{et~al.}(2022)Li, Yu, Yuan, and Chen]{YuanLPR2022}
Li,~G.; Yu,~D.; Yuan,~L.; Chen,~X. Single Pulse Manipulations in Synthetic
  Time-Frequency Space. \emph{Laser \& Photonics Rev.} \textbf{2022},
  \emph{16}, 2100340\relax
\mciteBstWouldAddEndPuncttrue
\mciteSetBstMidEndSepPunct{\mcitedefaultmidpunct}
{\mcitedefaultendpunct}{\mcitedefaultseppunct}\relax
\EndOfBibitem
\bibitem[Kielpinski \latin{et~al.}(2011)Kielpinski, Corney, and
  Wiseman]{PRL2011WisemanShapeConversion}
Kielpinski,~D.; Corney,~J.~F.; Wiseman,~H.~M. Quantum Optical Waveform
  Conversion. \emph{Phys. Rev. Lett.} \textbf{2011}, \emph{106}, 130501\relax
\mciteBstWouldAddEndPuncttrue
\mciteSetBstMidEndSepPunct{\mcitedefaultmidpunct}
{\mcitedefaultendpunct}{\mcitedefaultseppunct}\relax
\EndOfBibitem
\bibitem[Chen \latin{et~al.}(2016)Chen, Taylor, and
  Yu]{NanfangRepProgPhys2016Review}
Chen,~H.-T.; Taylor,~A.~J.; Yu,~N. A review of metasurfaces: physics and
  applications. \emph{Rep. Prog. Phys.} \textbf{2016}, \emph{79}, 076401\relax
\mciteBstWouldAddEndPuncttrue
\mciteSetBstMidEndSepPunct{\mcitedefaultmidpunct}
{\mcitedefaultendpunct}{\mcitedefaultseppunct}\relax
\EndOfBibitem
\bibitem[Qiu \latin{et~al.}(2021)Qiu, Zhang, Hu, and Kivshar]{QiuNanoLett2021}
Qiu,~C.-W.; Zhang,~T.; Hu,~G.; Kivshar,~Y. Quo Vadis, Metasurfaces? \emph{Nano
  Lett.} \textbf{2021}, \emph{21}, 5461--5474\relax
\mciteBstWouldAddEndPuncttrue
\mciteSetBstMidEndSepPunct{\mcitedefaultmidpunct}
{\mcitedefaultendpunct}{\mcitedefaultseppunct}\relax
\EndOfBibitem
\bibitem[Dorrah and Capasso(2022)Dorrah, and Capasso]{CapassoScience2022}
Dorrah,~A.~H.; Capasso,~F. Tunable structured light with flat optics.
  \emph{Science} \textbf{2022}, \emph{376}, eabi6860\relax
\mciteBstWouldAddEndPuncttrue
\mciteSetBstMidEndSepPunct{\mcitedefaultmidpunct}
{\mcitedefaultendpunct}{\mcitedefaultseppunct}\relax
\EndOfBibitem
\bibitem[Divitt \latin{et~al.}(2019)Divitt, Zhu, Zhang, Lezec, and
  Agrawal]{DivittScience2019MetaFS}
Divitt,~S.; Zhu,~W.; Zhang,~C.; Lezec,~H.~J.; Agrawal,~A. Ultrafast optical
  pulse shaping using dielectric metasurfaces. \emph{Science} \textbf{2019},
  \emph{364}, 890--894\relax
\mciteBstWouldAddEndPuncttrue
\mciteSetBstMidEndSepPunct{\mcitedefaultmidpunct}
{\mcitedefaultendpunct}{\mcitedefaultseppunct}\relax
\EndOfBibitem
\bibitem[Abbaszadeh \latin{et~al.}(2021)Abbaszadeh, Tehranian, and
  Salehi]{SalehiOE2021}
Abbaszadeh,~A.; Tehranian,~A.; Salehi,~J.~A. Phase-only femtosecond optical
  pulse shaping based on an all-dielectric polarization-insensitive
  metasurface. \emph{Opt. Express} \textbf{2021}, \emph{29}, 36900--36914\relax
\mciteBstWouldAddEndPuncttrue
\mciteSetBstMidEndSepPunct{\mcitedefaultmidpunct}
{\mcitedefaultendpunct}{\mcitedefaultseppunct}\relax
\EndOfBibitem
\bibitem[Geromel \latin{et~al.}(2023)Geromel, Georgi, Protte, Lei, Bartley,
  Huang, and Zentgraf]{ThomasNanoLett2023}
Geromel,~R.; Georgi,~P.; Protte,~M.; Lei,~S.; Bartley,~T.; Huang,~L.;
  Zentgraf,~T. Compact Metasurface-Based Optical Pulse-Shaping Device.
  \emph{Nano Lett.} \textbf{2023}, \emph{23}, 3196--3201\relax
\mciteBstWouldAddEndPuncttrue
\mciteSetBstMidEndSepPunct{\mcitedefaultmidpunct}
{\mcitedefaultendpunct}{\mcitedefaultseppunct}\relax
\EndOfBibitem
\bibitem[Salary and Mosallaei(2020)Salary, and Mosallaei]{HosseinACSPho2020MS}
Salary,~M.~M.; Mosallaei,~H. Tunable All-Dielectric Metasurfaces for Phase-Only
  Modulation of Transmitted Light Based on Quasi-bound States in the Continuum.
  \emph{ACS Photonics} \textbf{2020}, \emph{7}, 1813--1829\relax
\mciteBstWouldAddEndPuncttrue
\mciteSetBstMidEndSepPunct{\mcitedefaultmidpunct}
{\mcitedefaultendpunct}{\mcitedefaultseppunct}\relax
\EndOfBibitem
\bibitem[Mohammadi~Dinani and Mosallaei(2023)Mohammadi~Dinani, and
  Mosallaei]{HosseinAPR2023}
Mohammadi~Dinani,~H.; Mosallaei,~H. Active Tunable Pulse Shaping Using
  MoS2-Assisted All-Dielectric Metasurface. \emph{Adv. Photo. Res.}
  \textbf{2023}, \emph{4}, 2200207\relax
\mciteBstWouldAddEndPuncttrue
\mciteSetBstMidEndSepPunct{\mcitedefaultmidpunct}
{\mcitedefaultendpunct}{\mcitedefaultseppunct}\relax
\EndOfBibitem
\bibitem[Liu \latin{et~al.}(2024)Liu, Liu, Tian, Tian, Li, Ren, Gong, and
  Gu]{QiLiuOE2024}
Liu,~Q.; Liu,~X.; Tian,~Y.; Tian,~Z.; Li,~G.; Ren,~X.-F.; Gong,~Q.; Gu,~Y.
  Parallel beam splitting based on gradient metasurface: from classical to
  quantum. \emph{Opt. Express} \textbf{2024}, \emph{32}, 31389--31404\relax
\mciteBstWouldAddEndPuncttrue
\mciteSetBstMidEndSepPunct{\mcitedefaultmidpunct}
{\mcitedefaultendpunct}{\mcitedefaultseppunct}\relax
\EndOfBibitem
\bibitem[Tian \latin{et~al.}(2025)Tian, Liu, Tian, and Gu]{TianSupp}
Tian,~Z.; Liu,~Q.; Tian,~Y.; Gu,~Y. Supp. \emph{See Supporting Information}
  \textbf{2025}, \relax
\mciteBstWouldAddEndPunctfalse
\mciteSetBstMidEndSepPunct{\mcitedefaultmidpunct}
{}{\mcitedefaultseppunct}\relax
\EndOfBibitem
\bibitem[Fearn and Loudon(1989)Fearn, and Loudon]{LoudonJoAB1989}
Fearn,~H.; Loudon,~R. Theory of two-photon interference. \emph{J. Opt. Soc. Am.
  B} \textbf{1989}, \emph{6}, 917--927\relax
\mciteBstWouldAddEndPuncttrue
\mciteSetBstMidEndSepPunct{\mcitedefaultmidpunct}
{\mcitedefaultendpunct}{\mcitedefaultseppunct}\relax
\EndOfBibitem
\bibitem[Tillmann \latin{et~al.}(2015)Tillmann, Tan, Stoeckl, Sanders,
  de~Guise, Heilmann, Nolte, Szameit, and Walther]{PRX2015QuanInterPhilip}
Tillmann,~M.; Tan,~S.-H.; Stoeckl,~S.~E.; Sanders,~B.~C.; de~Guise,~H.;
  Heilmann,~R.; Nolte,~S.; Szameit,~A.; Walther,~P. Generalized Multiphoton
  Quantum Interference. \emph{Phys. Rev. X} \textbf{2015}, \emph{5},
  041015\relax
\mciteBstWouldAddEndPuncttrue
\mciteSetBstMidEndSepPunct{\mcitedefaultmidpunct}
{\mcitedefaultendpunct}{\mcitedefaultseppunct}\relax
\EndOfBibitem
\bibitem[Ferreri \latin{et~al.}(2019)Ferreri, Ansari, Silberhorn, and
  Sharapova]{PRA2019HOMSilberhorn}
Ferreri,~A.; Ansari,~V.; Silberhorn,~C.; Sharapova,~P.~R. Temporally multimode
  four-photon Hong-Ou-Mandel interference. \emph{Phys. Rev. A} \textbf{2019},
  \emph{100}, 053829\relax
\mciteBstWouldAddEndPuncttrue
\mciteSetBstMidEndSepPunct{\mcitedefaultmidpunct}
{\mcitedefaultendpunct}{\mcitedefaultseppunct}\relax
\EndOfBibitem
\bibitem[Tian \latin{et~al.}(2024)Tian, Liu, Tian, and Gu]{JOSABTZH2024}
Tian,~Z.; Liu,~Q.; Tian,~Y.; Gu,~Y. Wavepacket interference of two photons
  through a beam splitter: from temporal entanglement to wavepacket shaping.
  \emph{J. Opt. Soc. Am. B} \textbf{2024}, \emph{41}, 2668--2674\relax
\mciteBstWouldAddEndPuncttrue
\mciteSetBstMidEndSepPunct{\mcitedefaultmidpunct}
{\mcitedefaultendpunct}{\mcitedefaultseppunct}\relax
\EndOfBibitem
\bibitem[Stobi{\'{n}}ska \latin{et~al.}(2009)Stobi{\'{n}}ska, Alber, and
  Leuchs]{LeuchsPerfectExcitFreeSpace}
Stobi{\'{n}}ska,~M.; Alber,~G.; Leuchs,~G. Perfect excitation of a matter qubit
  by a single photon in free space. \emph{EPL} \textbf{2009}, \emph{86},
  14007\relax
\mciteBstWouldAddEndPuncttrue
\mciteSetBstMidEndSepPunct{\mcitedefaultmidpunct}
{\mcitedefaultendpunct}{\mcitedefaultseppunct}\relax
\EndOfBibitem
\bibitem[Jin \latin{et~al.}(2023)Jin, Zhang, Liu, Liang, Liu, Hu, Li, Wang,
  Yang, Zhu, and Li]{GuixinNanoLett2023}
Jin,~M.; Zhang,~X.; Liu,~X.; Liang,~C.; Liu,~J.; Hu,~Z.; Li,~K.; Wang,~G.;
  Yang,~J.; Zhu,~L.; Li,~G. A Centimeter-Scale Dielectric Metasurface for the
  Generation of Cold Atoms. \emph{Nano Lett.} \textbf{2023}, \emph{23},
  4008--4013\relax
\mciteBstWouldAddEndPuncttrue
\mciteSetBstMidEndSepPunct{\mcitedefaultmidpunct}
{\mcitedefaultendpunct}{\mcitedefaultseppunct}\relax
\EndOfBibitem
\bibitem[Balthasar~Mueller \latin{et~al.}(2017)Balthasar~Mueller, Rubin,
  Devlin, Groever, and Capasso]{PRL2017Capasso}
Balthasar~Mueller,~J.~P.; Rubin,~N.~A.; Devlin,~R.~C.; Groever,~B.; Capasso,~F.
  Metasurface Polarization Optics: Independent Phase Control of Arbitrary
  Orthogonal States of Polarization. \emph{Phys. Rev. Lett.} \textbf{2017},
  \emph{118}, 113901\relax
\mciteBstWouldAddEndPuncttrue
\mciteSetBstMidEndSepPunct{\mcitedefaultmidpunct}
{\mcitedefaultendpunct}{\mcitedefaultseppunct}\relax
\EndOfBibitem
\bibitem[Wang \latin{et~al.}(2018)Wang, Jing, Sun, Yang, Yu, Tamma, Bao, and
  Pan]{PanPRL2018}
Wang,~X.-J.; Jing,~B.; Sun,~P.-F.; Yang,~C.-W.; Yu,~Y.; Tamma,~V.; Bao,~X.-H.;
  Pan,~J.-W. Experimental Time-Resolved Interference with Multiple Photons of
  Different Colors. \emph{Phys. Rev. Lett.} \textbf{2018}, \emph{121},
  080501\relax
\mciteBstWouldAddEndPuncttrue
\mciteSetBstMidEndSepPunct{\mcitedefaultmidpunct}
{\mcitedefaultendpunct}{\mcitedefaultseppunct}\relax
\EndOfBibitem
\bibitem[Yard \latin{et~al.}(2024)Yard, Jones, Paesani, Ma\"{\i}nos, Bulmer,
  and Laing]{AnthonyPRL2024}
Yard,~P.; Jones,~A.~E.; Paesani,~S.; Ma\"{\i}nos,~A.; Bulmer,~J. F.~F.;
  Laing,~A. On-Chip Quantum Information Processing with Distinguishable
  Photons. \emph{Phys. Rev. Lett.} \textbf{2024}, \emph{132}, 150602\relax
\mciteBstWouldAddEndPuncttrue
\mciteSetBstMidEndSepPunct{\mcitedefaultmidpunct}
{\mcitedefaultendpunct}{\mcitedefaultseppunct}\relax
\EndOfBibitem
\end{mcitethebibliography}

\providecommand{\latin}[1]{#1}
\makeatletter
\providecommand{\doi}
  {\begingroup\let\do\@makeother\dospecials
  \catcode`\{=1 \catcode`\}=2 \doi@aux}
\providecommand{\doi@aux}[1]{\endgroup\texttt{#1}}
\makeatother
\providecommand*\mcitethebibliography{\thebibliography}
\csname @ifundefined\endcsname{endmcitethebibliography}
  {\let\endmcitethebibliography\endthebibliography}{}

\end{document}